# Enhanced Thermal Transport across the Interface between Charged Graphene Electrodes and Poly(ethylene oxide) Electrolytes by Non-covalent Functionalization

*Siyu Tian, [a,‡] Dezhao Huang, [b,‡] Zhihao Xu, [b] Shiwen Wu, [a] Tengfei Luo, [b,*] Guoping Xiong [a,*]*

**ABSTRACT:** Interfacial thermal transport between electrodes and polymer electrolytes can play a crucial role in the thermal management of solid-state lithium-ion batteries (SLIBs). Modifying the electrode surface with functional molecules can effectively increase the interfacial thermal conductance (ITC) between electrodes and polymers (e.g., electrolytes, separators); however, how they influence the interfacial thermal transport in SLIBs during charge/discharge remains unknown. In this work, we conduct molecular dynamics (MD) simulations to investigate the ITC between charged electrodes and solid-state polymer electrolytes (SPEs) mixed with ionic liquids (ILs). We find that ILs could self assemble at the electrode surface and act as non-covalent functional molecules that could significantly enhance the interfacial thermal transport during charge/discharge because of the formation of a densely packed cationic or anionic layer at the interface. While the electrostatic interactions between the charged electrode and the IL ions are responsible for forming these dense interfacial layers, the enhancement of ITC is mainly contributed by the increased Lennard-Jones (LJ) interactions between the charged electrodes and ILs. This work may provide useful insights into the understanding of interfacial thermal transport between electrodes and electrolytes of SLIBs during charge/discharge.

## 1. INTRODUCTION

Solid-state lithium-ion batteries (SLIBs) with high energy density, power density, and reliability are desired in many applications such as electric vehicles and portable electronics. Recent development in new nanomaterials has significantly improved the electrochemical performance of SLIBs [1, 2]. However, these high-performance SLIBs inevitably generate a large amount of heat during charge/discharge, particularly at high rates, within a limited cell space and still suffer from inefficient heat dissipation because of their relatively low cell thermal conductivity (0.2 – 0.6 W m$^{-1}$ K$^{-1}$) [3, 4]. Without efficient thermal management, the generated heat in SLIBs can result in many issues, including considerable temperature rise (i.e., overheating), performance degradation, and even catastrophic failure (e.g., thermal runaway) of the batteries [5]. Compared with external thermal management strategies such as using passive cooling systems, enhancing thermal transport such as that at the interfaces between electrodes and solid-state polymer electrolytes (SPEs) within batteries can contribute to addressing these issues.

Understanding fundamental physics is vital to achieving efficient thermal transport of SLIBs. A SLIB contains current collectors, cathode, anode, and SPEs mixed with lithium salts. Previous studies have shown that the thermal resistance of SLIBs is mainly contributed by the low-thermal-conductivity SPEs and the SPE/electrode interfaces [6-11]. For achieving enhanced interfacial thermal transport between electrode/polymer interfaces, various functionalization approaches have been investigated. For instance, He et al. evaluated the interfacial thermal transport between poly(ethylene) oxide (PEO) and cathode by conducting molecular dynamics (MD) simulations [12]. They found that the interfacial thermal conductance (ITC) was greatly enhanced by ~200% when functionalizing the cathode surface with a layer of poly(acrylic acid) (PAA). Further calculations showed that the enhancement of ITC mainly originated from the formation of strong hydrogen bonds between PAA and PEO molecules. Dhakane et al. studied the interfacial thermal transport between polyethylene (PE) and cathode [13]. The ITC was increased by 250% when functionalizing the electrode surface with 3-Aminopropyl triethoxysilane, which correlated well with their experimental results. Although the enhancement of ITC between the electrodes and polymers using these functional molecules is encouraging, their impact on electrochemical performance (e.g., ionic conductivity and stability) of SLIBs must be further testified. Indeed, the addition of PAA resulted in significant Coulombic and capacity losses of LIBs because of the parasitic electrochemical reactions of PAA molecules, as shown in previous experimental studies [14]. Consequently, finding functional molecules that are compatible with electrode materials and SPEs to enhance interfacial thermal transport is important to balance the electrochemical and thermal performance of SLIBs.

In contrast to the foregoing functional molecules, ionic liquids (ILs) consisting of self-dissociable cation-anion pairs are thermally and electrochemically stable. Previous studies have shown that ILs can effectively increase the ionic conductivity of SPEs owing to their plasticization effect [15-17]. Moreover, experimental reports have shown that imidazolium ILs as non-covalent functionalization molecules can enhance the interfacial thermal transport between graphene and polymers [18-20]. However, the mechanisms of such IL-induced enhancement of thermal transport in graphene/polymer composites remain unknown. Furthermore, electric charges accumulate on the electrode surface during charge/discharge processes of SLIBs, resulting in strong Coulombic interactions between electrodes and ILs, which may lead to different interfacial thermal transport mechanisms. Understanding the influence of ILs on the interfacial thermal transport between charged electrodes and SPEs is thus intriguing and beneficial to the development of safe, high-performance SLIBs. To date, the effect of electrode charge state on the ITC of SLIBs has not ever been reported in prior work, and the fundamental mechanisms warrant a systematic study.

Here, we conduct systematic MD simulations to investigate the influence of imidazolium ILs on interfacial thermal transport between charged graphene electrodes and amorphous PEO electrolytes during charge/discharge of SLIBs. The results indicate that ITC increases as the density of charge on graphene surface increases. Furthermore, ITC values are higher when graphene surface is negatively charged than when it is positively charged with the same charge density. By analyzing the vibrational spectra of interfacial species, we find that imidazolium cations exhibit a better coupling effect between graphene and PEO, which agrees well with the ITC results. The Coulombic interaction between the charged graphene and IL ions plays a key role in attracting the IL molecules to the interface. Interestingly, the decomposition of thermal conductance shows that the Lennard-Jones (LJ) interactions between charged electrodes and IL ions are the main contributor to ITC. This study can potentially provide new insights to design high-performance SPEs with improved thermal transport properties for practical energy storage applications.

## 2. SIMULATION METHODS

The simulation system consists of 30 P(EO)$_{50}$ chains and 120 1-ethyl-3-methylimidazolium tetrafluoroborate ([EMIM][BF$_4$]) molecules at each side of the single-layer graphene (Fig. 1a). The in-plane size of graphene is ~ 43 × 45 nm$^2$ (720 carbon atoms) with charge densities ranging from -0.594 to 0.594 C m$^{-2}$ to represent the charged and discharged states of electrodes. Graphene charge density is varied by assigning a constant charge to each carbon atom. To keep the electric neutrality of the simulated system, we change the number of anions and cations of [EMIM][BF$_4$] according to the assigned graphene charges, which are summarized in Table S1. The Tersoff potential is employed to describe the interactions between the carbon atoms in graphene. The all-atom optimized potentials for liquid simulation (OPLS-AA) force field is used to model the PEO matrix. All the parameters of [EMIM][BF$_4$] are adapted from the revised OPLS-2009IL force field [21, 22]. A cutoff distance of 10 Å is used for all non-bonding interactions. The 12-6 LJ parameters between graphene and IL-PEO mixture are calculated by Lorentz-Berthelot mixing rules ($\epsilon_{ij}$ = sqrt($\epsilon_i \epsilon_j$); $\sigma_{ij} = (\sigma_i + \sigma_j)/2$), where $\epsilon$ and σ are the energy and distance constants, respectively. The long-range Coulombic interaction is evaluated by the particle-particle-particle-mesh (PPPM) algorithm with an accuracy of $1 \times 10^{-4}$ [23]. A time step of 0.25 fs is used for all simulations because of the light-weight hydrogen atoms [24, 25].

All simulations are conducted using the Large-scale Atomic/Molecular Massively Parallel Simulator (LAMMPS) [26]. The interfacial thermal transport between the graphene and PEO matrix in the presence of ILs is evaluated using the non-equilibrium molecular dynamics (NEMD) simulation at 300 K and 1 atm. The simulation system is first relaxed in the isothermal-isobaric ensemble (NPT) at 300 K for 1 ns, followed by annealing from 300 to 600 K in the NPT with a heating rate of 50 K/ns. The annealed system is further equilibrated at 600 K for 12 ns, enabling the homogeneous distribution of ILs in the PEO matrix. Finally, the system is quenched to 300 K and equilibrated for 2 ns in the NPT ensemble to converge the density. A final 2 ns of relaxation in the canonical ensemble (NVT) at 300 K is performed before the 4 ns of microcanonical ensemble (NVE) NEMD production simulation. In the NEMD simulation, the heat source and heat sink are maintained at 350 K and 250 K using Langevin thermostat, respectively. The two boundaries in the z-direction are stabilized by two 3 Å-thick fixed layers, which are excluded from all calculations. Two 5 Å-thick vacuum layers next to these two fixed layers are added to prevent heat leakage. The ITC with standard deviations is calculated using the last 2 ns of the production period. The same simulation procedure is applied to other systems with different graphene charge densities ranging from -0.594 to 0.594 C m$^{-2}$. Figure 1a shows the representative simulation structure of graphene (0 C m$^{-2}$) and IL-PEO mixture at equilibrium state as the baseline case. The corresponding steady-state temperature profile of the simulation system is shown in Figure 1b. At the steady-state of NEMD simulation, the ITC is calculated using $G = q/\Delta T$, where $G$ is the ITC, $q$ is the heat flux, and $\Delta T$ is the temperature drop across the two graphene/IL-PEO interfaces.

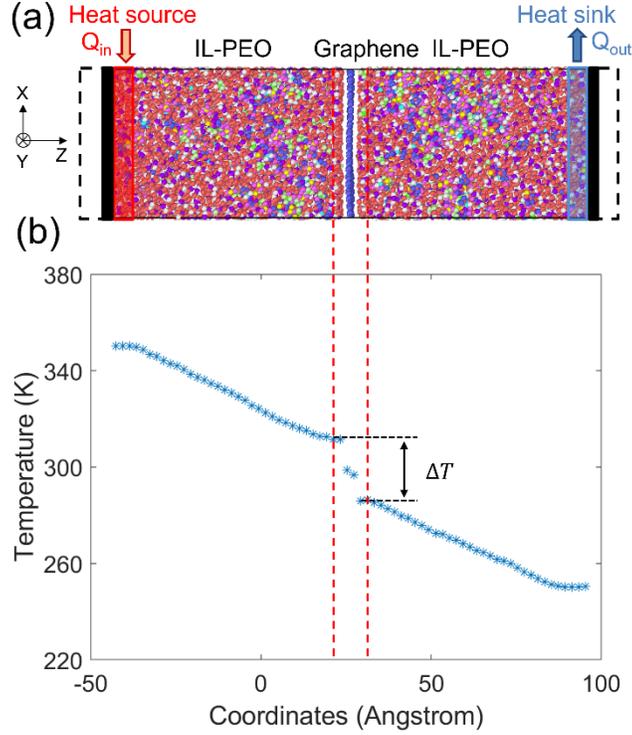

Figure 1. (a) Simulation setup of the baseline case showing the interface structure between graphene (0 C m$^{-2}$) and IL-PEO. ITC is calculated via NEMD: heat source (red) and heat sink (blue) are maintained at 350 K and 250 K using Langevin thermostat, respectively. The two boundaries (black) in the z-direction are fixed by 3 Å and excluded from all calculations. Two 5 Å-thick vacuum layers (areas with black dashed lines) next to these two fixed layers are added to prevent heat leakage. (b) The temperature profile of the simulation system with a graphene charge density of 0 C m$^{-2}$, showing the interfacial temperature difference ($\Delta T$).

## 3. RESULTS AND DISCUSSION

### 3.1 Force field validation

Thermal conductivities of PEO and IL-PEO mixture are calculated using the same procedure described above to validate MD calculations. First, the density and thermal conductivity of [EMIM][BF$_4$] are calculated to be 1.21 g cm$^{-3}$ and 0.23 W m$^{-2}$ K$^{-1}$, respectively, close to the corresponding experimental values of 1.28 g cm$^{-3}$ and 0.2 W m$^{-2}$ K$^{-1}$ [21, 22, 27]. The mass density of amorphous PEO at equilibrium (300 K, 1 atm) is calculated to be 1.08 g cm$^{-3}$. The thermal conductivity of PEO is found to be 0.28 W m$^{-2}$ K$^{-1}$. Both are in good agreement with experimental results, with corresponding values of ~1.11 g cm$^{-3}$ and 0.2 to 0.37 W m$^{-2}$ K$^{-1}$ [9, 28-30]. The calculated ITCs of graphene/PEO and graphene/IL-PEO are 61 $\pm$ 2 MW m$^{-2}$ K$^{-1}$ and 62 $\pm$ 5 MW m$^{-2}$ K$^{-1}$, respectively. Although no data are reported for graphene/PEO and graphene/IL-PEO interfaces, the ITC values calculated in this work are comparable with those of the graphene/polymer interfaces [24, 31-33].

### 3.2 Charge-dependent ITC

During cyclic charge/discharge processes, the graphene electrode becomes either negatively or positively charged. Ions in the IL (i.e., [EMIM][BF$_4$]) will diffuse towards the charged electrode surface and form a compact interfacial layer because of the strong Coulombic interactions between the electrode and ions. To simulate the relevant charge/discharge states, we vary the surface charge densities of the graphene electrode to evaluate their impact on the interfacial thermal transport between the electrode and polymer electrolyte. Figure 2 shows the calculated ITC of the graphene/IL-PEO system with different graphene surface charge densities ranging from -0.594 to 0.594 C m$^{-2}$. The results show that both positive and negative charges on the graphene electrode surface can enhance the ITC. When the graphene charge density varies from 0 to -0.594 C m$^{-2}$, the ITC increases from 62 $\pm$ 5 MW m$^{-2}$ K$^{-1}$ to 277 $\pm$ 55 MW m$^{-2}$ K$^{-1}$ (447%). Similarly, when the graphene charge density varies from 0 to 0.594 C m$^{-2}$, the ITC increases from 62 $\pm$ 5 MW m$^{-2}$ K$^{-1}$ to 180 $\pm$ 26 MW m$^{-2}$ K$^{-1}$ (290%). For comparison, ITC values of the graphene/PEO system without the IL are calculated to be 159 $\pm$ 17 MW m$^{-2}$ K$^{-1}$ and 156 $\pm$ 10 MW m$^{-2}$ K$^{-1}$ when the graphene charge density is -0.594 C m$^{-2}$ and 0.594 C m$^{-2}$, respectively. Furthermore, the ITC values corresponding to the negatively charged graphene (e.g., 277 ± 55 MW m$^{-2}$ K$^{-1}$ at -0.594 C m$^{-2}$) are noticeably higher than those corresponding to the positively charged graphene at the same charge density (e.g., 180 ± 26 MW m$^{-2}$ K$^{-1}$ at 0.594 C m$^{-2}$). These results indicate that the interfacial polymer-IL-electrode structure is significantly affected by the electric charges on the electrode surface during charge/discharge processes, illustrating the importance of ILs in promoting efficient thermal transport across the graphene/polymer interface.

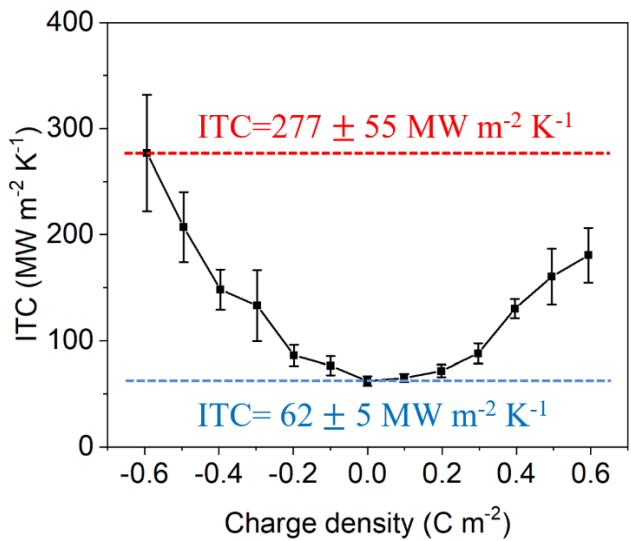

Figure 2. Calculated ITC values as a function of the graphene charge density. The red and blue dashed lines represent the highest and lowest ITC values of graphene/IL-PEO with a charge density of -0.594 C m$^{-2}$ and 0 C m$^{-2}$, respectively.

To understand the mechanisms of charge-dependent enhancement of ITC, atomic structures and interfacial structures of the graphene/IL-PEO system at different charge densities are investigated. The IL is evenly distributed in the PEO matrix when graphene is uncharged and can barely be observed at the graphene/IL-PEO interface (Figure 3a). When the graphene surface undergoes charge and discharge, the counterions from the IL diffuse toward the charged graphene surface. Consequently, an intermediated and densely-packed EMIM$^+$ (Figure 3b) or BF$_4^-$ layer (Figure 3c) is formed at the interface because of the strong Coulombic interaction between the ions in the IL and the charged graphene, a phenomenon similar to the formation of an electric double layer of ILs in solid-liquid systems [34-37]. To quantitively evaluate the interfacial structures, the density of the IL and radial distribution function (RDF, defined as $g(r)$) of the IL with respect to the graphene carbon atoms are calculated (Figure S1). The first peak densities of the $EMIM^+$ and $BF_4^-$ ions near the graphene surface are summarized in Figure 3d. The peak density of cations increases from 0.22 to 1.76 g cm$^{-3}$ when the charge density changes from 0 to -0.594 C m$^{-2}$. As shown in Figure S1a, the interfacial species is dominated by $EMIM^+$ with a graphene surface charge density of -0.594 C m$^{-2}$. On the other hand, when the charge density changes from 0 to 0.594 C m$^{-2}$, the peak density of anions near the graphene surface increases from 0.18 g cm$^{-3}$ to 1.85 g cm$^{-3}$ (Figure 3d). Similarly, most of the ions near the positively charged graphene surface (0.594 C m$^{-2}$) are $BF_4^-$ (Figure S1c). The location of the first peak of $g(r)$ is summarized in Figure 3e. Although more IL ions are accumulated at the interface with increased graphene charge densities, the distance between carbon atoms in graphene and ions in the IL remains relatively stable. For instance, the C-N (corresponding to negatively charged graphene) and C-B (corresponding to positively charged graphene) distances are calculated to be ~4.7 Å and ~4.3 Å, respectively. These findings suggest that the long-range Coulombic interaction due to charged graphene will draw more IL ions to the surface, but the equilibrium inter-molecular distance is not altered, which should be dominated by the stronger but shorter range LJ interaction. The similar trends of the interfacial structures of graphene/IL-PEO and ITC values at different charge states suggest that the intermediated IL layer and its density are key to the enhanced thermal conductance across the graphene/IL-PEO interface. Although both $EMIM^+$ and $BF_4^-$ have similar pack densities at the interface (Figure 3d) and comparable inter-molecular distances with respect to graphene surface (Figure 3e), $EMIM^+$ performs better than $BF_4^-$ in enhancing the interfacial thermal transport between graphene and PEO, which may be attributed to their phonon spectral features as discussed in the next section.

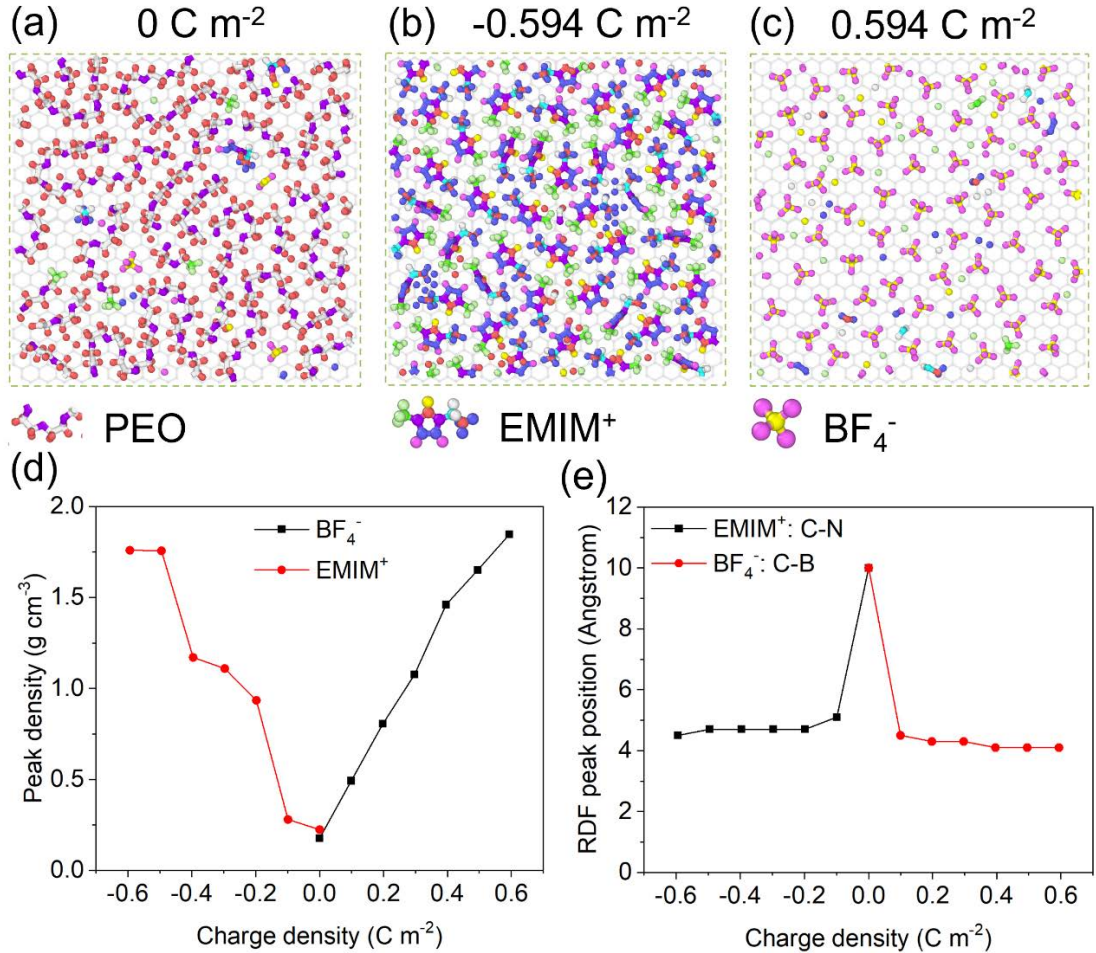

Figure 3. Atomic structure of the graphene/IL-PEO system with different charge densities for (a) uncharged graphene (0 C m$^{-2}$), (b) negatively charged graphene (-0.594 C m$^{-2}$), and (c) positively charged graphene (0.594 C m$^{-2}$), showing the accumulation of cations (EMIM$^+$) and anions (BF$_4^-$) at negatively and positively charged graphene surfaces, respectively. (d) The first peak densities of EMIM$^+$ and BF$_4^-$ near the graphene surfaces with various charge densities. (e) The locations of the first RDF peak of graphene carbon atoms with respect to the nitrogen atoms in EMIM$^+$ (C-N) and boron atoms in BF$_4^-$ (C-B) corresponding to negatively and positively charged graphene, respectively.

### 3.3 Vibrational density of states (VDOS)

To further understand the mechanisms of the enhanced ITC between charged graphene and IL-PEO, we performed vibrational spectra analysis to evaluate the vibrational coupling across the interface. In the graphene/IL-PEO system, particular attention is directed to understand the role of IL ions that dominate the interfacial structures when the graphene surface is charged. The VDOS is calculated by performing Fourier transform (FT) of the velocity autocorrelation function (VACF) of atoms through equation 1 [38, 39]:

$$D(\omega) = \int_0^\tau \Gamma(t) \cos(\omega t) \, dt \tag{1}$$

where $\omega$ is frequency, $D(\omega)$ is the vibration power spectrum at frequency $\omega$, and $\Gamma(t)$ is the atomic VACF defined by equation 2:

$$\Gamma(t) = <v(t)v(0)> \tag{2}$$

where $v(0)$ and $v(t)$ are the velocities of the atoms at time 0 and $t$, respectively.

Figure 4 shows the calculated VDOS of each component in the graphene/IL-PEO system with a graphene charge density of -0.594 C m$^{-2}$. The VDOS of the IL is decomposed to the spectra of EMIM$^+$ and BF$_4^-$. The spectra of graphene agrees well with the results reported in prior work using the Tersoff potential [40-43]. The VDOS of PEO shows several major peaks at 5 THz, 30 THz, 42 THz and 90 THz, which are in good agreement with the results reported by Meng et al. [44]. We observe that the EMIM$^+$ vibrational energy is more evenly distributed across a large frequency range from ~5 to 50 THz, while that of the BF$_4^-$ are more localized surrounding the above-mentioned few peaks. The overlap of VDOS between different interfacial species is known to be indicative of the thermal transport across solid/solid and solid/liquid interfaces due to elastic channels [45, 46]. By comparing the VDOS of the interfacial species, we find that BF$_4^-$ anions apparently exhibit less overlap (indicated by the green area) with graphene and PEO. Meanwhile, the VDOS of EMIM$^+$ has more significant overlap with other interfacial species, especially in the low- and middle-frequency ranges, in which most heat transfer occurs. Consequently, high thermal conductance is obtained at the interface when graphene is negatively charged, which agrees well with the ITC values in Figure 2 and is consistent with prior results [34]. It is worth noting that when EMIM$^+$ forms a layer at the interface, it can work as an effective "vibrational bridge" that bridges the vibrational mismatch between graphene and PEO and thus enhance the effective ITC [47, 48]. BF$_4^-$ can play a similar role when form an interfacial layer, but it is less effective than EMIM$^+$ due to its comparatively less vibrational spectral overlap with graphene and PEO.

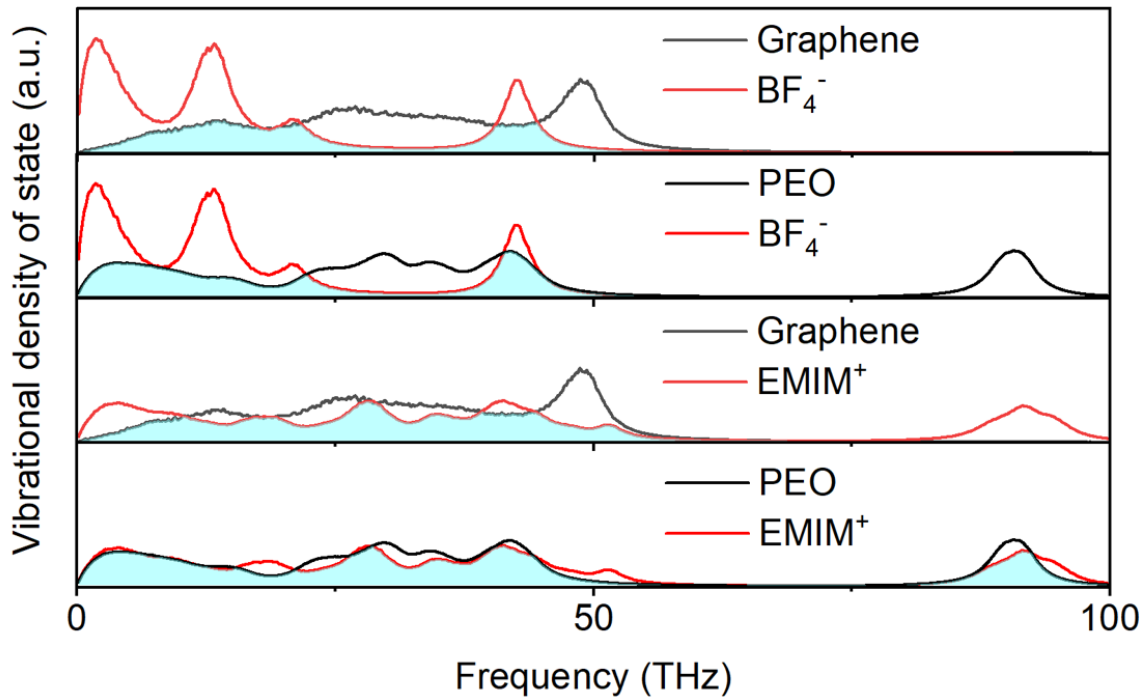

Figure 4. Normalized VDOS of graphene, EMIM$^+$, BF$_4^-$, and PEO in graphene/IL-PEO system with a charge density of -0.594 C m$^{-2}$. The overlap area is highlighted in light green.

### 3.4 Decomposition of ITC

In the graphene/IL-PEO system, the thermal energy transport across the interface is achieved through graphene-IL and graphene-PEO interactions. At different graphene charge densities, the contributions of graphene-IL and graphene-PEO interactions are expected to vary because the interfacial structure changes during charge/discharge processes. To further reveal the underlying mechanism, we quantify the contributions of graphene-IL and graphene-PEO interactions to ITC according to our prior work [25]. As shown in Figure 5a, we decompose the total ITC into contributions from graphene-IL and graphene-PEO interactions. The results show that the interfacial thermal energy is solely attributed to graphene-PEO interaction when the graphene is not charged. However, the thermal energy transferred through graphene-IL interaction significantly increases and dominates the interfacial heat transfer when the graphene electrode is in a charged state. We further decompose the heat flow and thus ITC into the LJ and Coulombic contributions (Figure 5b) according to $G^{LJ} = q^{LJ}/\Delta T$ or $G^Q = q^Q/\Delta T$, where LJ and Q denote the LJ and Coulombic interactions, respectively. The enhancement of ITC in the charged graphene/IL-PEO systems is mainly attributed to the increase of LJ contribution between the graphene electrode and IL, rather than that of the Coulombic interactions. Such an observation is consistent with our prior analysis [25, 49], where Coulombic interaction is responsible for attracting polar molecules closer to the interface, but LJ interaction is mainly responsible for thermal transport.

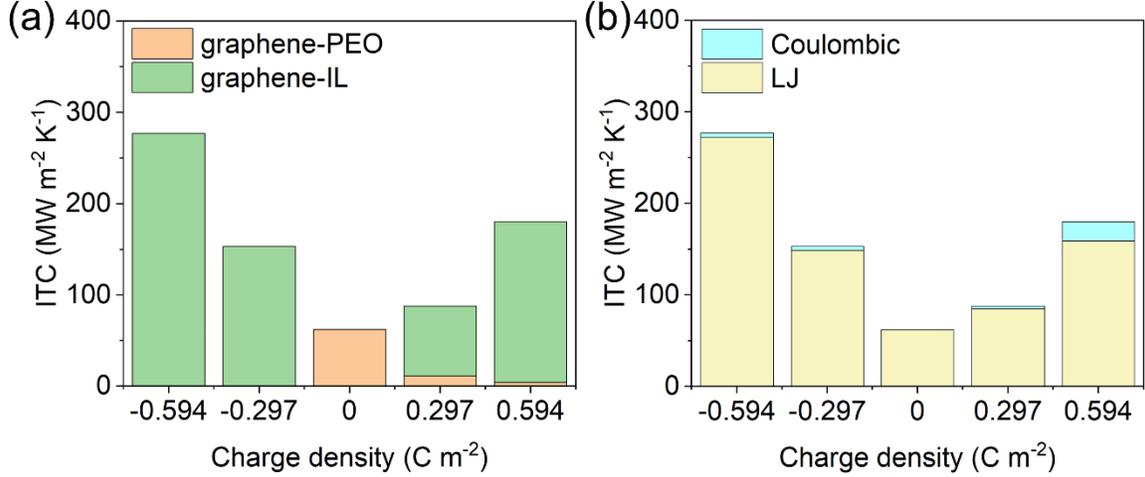

Figure 5. Decomposition of ITC into contributions from (a) graphene-PEO and graphene-IL interactions; and (b) Coulombic and LJ interactions.

## 4. CONCLUSIONS

In summary, we have calculated the ITC in the graphene/IL-PEO system with different graphene charge densities to understand the mechanisms of enhanced interfacial thermal transport between electrodes and SPEs in the charged and discharged states. We find that the ITC increases with the increase of graphene charge densities. The interfacial structures change accordingly with the graphene charge states, where IL ions form concentrated layers at the interface due to Coulombic attraction from the charged graphene. When the graphene is negatively or positively charged, the interfacial thermal transport is dominated by cations or anions of ILs, respectively. The ITC values of the system are higher when graphene is negatively charged than when it is positively charged, which can be well explained by analyzing the VDOS of interfacial species. By decomposing the total heat flux into contributions from graphene-PEO and graphene-IL interactions, we reveal that the interfacial thermal transport is dominated by the interactions between charged graphene and IL. Moreover, the decomposition of ITC shows that the enhanced interfacial thermal transport in the charged graphene/IL-PEO systems is primarily attributed to the enhancement of LJ interactions rather than the Coulombic interactions, with the later contribute indirectly by attracting IL ions to the interface. The results of this work will provide new insights into the understanding of interfacial thermal transport between electrodes and electrolytes of SLIBs during charge/discharge processes.

## ASSOCIATED CONTENT

**Supporting Information**

Nominal charge of graphene sheet, real charge per graphene carbon atom, graphene charge density and number of ions used in the simulations; the density distribution of cations ($EMIM^+$) and anions ($BF_4^-$) at the interface with various graphene charge densities (a) -0.594 C m$^{-2}$, (b) 0 C m$^{-2}$, and (c) 0.594 C m$^{-2}$; the radial distribution function of graphene carbon atoms with respect to the nitrogen atoms in $EMIM^+$ and boron atoms in $BF_4^-$ with various graphene charge densities (d) -0.594 C m$^{-2}$, (e) 0 C m$^{-2}$, and (f) 0.594 C m$^{-2}$; decomposition of interfacial thermal conductance (ITC) (PDF)

## AUTHOR INFORMATION

**Corresponding Authors**

Tengfei Luo − Chemical and Biomolecular Engineering, University of Notre Dame, Notre Dame, Indiana 46556, United States; Email: tluo@nd.edu

Guoping Xiong − Department of Mechanical Engineering, The University of Texas at Dallas, 800 W Campbell Rd, Richardson, TX 75080, United States; Email: guoping.xiong@utdallas.edu

**Authors**

Siyu Tian − Department of Mechanical Engineering, The University of Texas at Dallas, 800 W Campbell Rd, Richardson, TX 75080, United States;

Dezhao Huang − Chemical and Biomolecular Engineering, University of Notre Dame, Notre Dame, Indiana 46556, United States;

Zhihao Xu − Chemical and Biomolecular Engineering, University of Notre Dame, Notre Dame, Indiana 46556, United States;

Shiwen Wu − Department of Mechanical Engineering, The University of Texas at Dallas, 800 W Campbell Rd, Richardson, TX 75080, United States;

**Author Contributions**

S.T. and D.H contributed equally. G.X., T.L., D.H., and S.T. contributed to the study conception and design. S.T., D.H., Z.X., and S.W. performed the simulations and calculations. G.X. and T.L. supervised the simulations and calculations. All authors analyzed the data and interpreted the results. S.T. wrote the first draft of the manuscript. All the authors contributed to the writing of the manuscript and approved the final draft.

**Notes**

The authors declare no competing financial interest.

ACKNOWLEDGEMENT

G.X. acknowledges the financial support from The University of Texas at Dallas startup fund and NSF (1937949 and 1949962). T.L. acknowledges the financial support from the NSF (1937923, 1949910, 2001079 and 2040565). The simulations are supported by the computing resources provided by TACC Lonestar5 and Stampede.